# A Novel Tool to Evaluate the Accuracy of Predicting Survival in Cystic Fibrosis

**Corresponding author:**
Aasthaa Bansal
Pharmaceutical Outcomes Research and Policy Program, School of Pharmacy
University of Washington, Seattle, WA, USA
abansal@uw.edu

**Co-authors:**
Nicole Mayer-Hamblett
Departments of Pediatrics and Biostatistics, University of Washington, Seattle, WA
Seattle Children's Research Institute, Seattle WA, USA

Christopher H. Goss
Division of Pulmonary and Critical Care Medicine, Departments of Medicine and Pediatrics, University of Washington, Seattle WA, USA

Patrick J. Heagerty
Department of Biostatistics, University of Washington, Seattle WA, USA

**ABSTRACT**

**Background:** Effective allocation of limited donor lungs in cystic fibrosis (CF) requires accurate survival predictions, so that high-risk patients may be prioritized for transplantation. In practice, decisions about allocation are made dynamically, using routinely updated assessments. We present a novel tool for evaluating risk prediction models that, unlike traditional methods, captures the dynamic nature of decision-making.

**Methods:** Predicted risk is used as a score to rank incident deaths versus patients who survive, with the goal of ranking the deaths higher. The mean rank across deaths at a given time measures time-specific predictive accuracy; when assessed over time, it reflects time-varying accuracy.

**Results:** Applying this approach to CF Registry data on patients followed from 1993-2011, we show that traditional methods do not capture the performance of models used dynamically in the clinical setting. Previously proposed multivariate risk scores perform no better than forced expiratory volume in 1 second as a percentage of predicted normal ($FEV_1$%) alone. Despite its value for survival prediction, $FEV_1$% has a low sensitivity of 45% over time (for fixed specificity of 95%), leaving room for improvement in prediction. Finally, prediction accuracy with annually-updated $FEV_1$% shows minor differences compared to $FEV_1$% updated every 2 years, which may have clinical implications regarding the optimal frequency of updating clinical information.

**Conclusions:** It is imperative to continue to develop models that accurately predict survival in CF. Our proposed approach can serve as the basis for evaluating the predictive ability of these models by better accounting for their dynamic clinical use.

**Key words**: cystic fibrosis, lung transplantation, survival, risk prediction, accuracy

**INTRODUCTION**

Lung transplantation has been shown to improve survival for some cystic fibrosis (CF) patients whose disease is no longer amenable to more conventional medical therapies.[1,2] However, due to a shortage of donor lungs, a large number of wait-list patients die while awaiting transplantation employing the current allocation system in the US. In 2010-2012, the wait-list mortality rate was 15.4 per 100 wait-list years. Candidates aged 12-17 years had the highest wait-list mortality, at 19.7 deaths per 100 wait-list years, followed by those aged 18-34 years at approximately 18.5 deaths per 100 wait-list years.[1]

Despite an increase in the rate of lung transplants over the past decade, wait-list mortality rates continue to rise.[3] Accurate predictions of mortality are necessary so that limited donor lungs may be prioritized to patients who are at the greatest risk of death without transplantation. The goal is to use a patient's clinical characteristics to calculate the predicted risk of mortality within a specified time period and to rank or classify patients on the wait-list as those who are predicted to die soon versus those who are not.

Although forced expiratory volume in 1 second as a percentage of predicted normal ($FEV_1\%$) is a standard measure of pulmonary function used for assessing patient prognosis and recommending lung transplantation, its accuracy has been questioned.[4-6] Several risk prediction (or prognostic) models combining $FEV_1\%$ with other clinical factors have been proposed; however, it is unclear how accurate these models are, since their evaluation has failed to address classification accuracy and instead focused solely on measures of model fit or calibration. For example, Liou et al developed the most well known prediction model in CF that includes $FEV_1\%$ and several additional clinical features.[5] They assessed model fit, and concluded that their model was more accurate than $FEV_1\%$ alone for predicting 5-year mortality. However, only assessing how closely the model fits the observed data is not sufficient when the model is meant for classification.[7-9] While agreement between the predicted and actual numbers of deaths represents good model calibration, it does not necessarily translate to good classification accuracy.[10]

Two papers have evaluated predicted versus actual mortality, Aaron et al[11] and Mayer-Hamblett et al.[9] Aaron et al compared predicted versus actual numbers of deaths and concluded that their model was accurate for prediction of one-year survival. Mayer-Hamblett et al, however, addressed appropriate evaluation of their model using classification error rates, i.e. sensitivity and specificity, for classifying subjects based on 2-year mortality. They showed that their multivariate model did not perform better than $FEV_1\%$ alone, and that both are inadequate for use in practice. They concluded: "better clinical predictors of short-term mortality among patients with CF are needed."

In addition to classification error rates, appropriate evaluation of a risk model must account for time-varying measurements. In practice, key clinical parameters included in the lung allocation score (LAS) are updated at routine care visits and then used to guide decisions regarding listing status in the US. Interestingly, the utility of updating markers has never been empirically evaluated. We assess *time-varying* accuracy using appropriate time-dependent classification error rates. Such evaluation could also help assess how often patient information such as $FEV_1\%$ should be updated in practice before information becomes outdated and impacts accuracy.

It is imperative to continue to develop models that more accurately predict survival for CF patients. Prior studies that have proposed models have either failed to evaluate the accuracy of survival prediction, or have been limited to quantifying model accuracy according to an arbitrary baseline time point that does not reflect clinical practice where physicians update predictions as clinical parameters change. The primary objective of our work is to apply a novel statistical tool for evaluating the accuracy of survival models in CF, importantly - and unlike prior evaluations - accounting for the potentially time-varying predictions based on routinely updated clinical assessments. Application of the tool for evaluating existing published survival models in CF is presented, in addition to a demonstration of how the tool can be utilized to identify subgroups of CF patients for whom prediction of survival is more reliable.

**METHODS**

**Study cohort**

We used the Cystic Fibrosis Foundation (CFF) National Patient Registry (CFFPR) data, which consists of all patients who agreed to participate in the CF Registry and who were seen in a CFF-accredited care center in the US from 1986 through 2012. The CFFPR obtains written and informed consent from each participant. This analysis was approved by the CFFPR oversight committee and was granted Institutional Review Board Exempt Status by the University of Washington Human Subjects Division (due to being anonymized data).

We selected the prevalent subjects in 1993 as our study cohort to match the analysis of Liou et al.[5] Our study cohort consisted of 17,926 prevalent CF cases on January 1, 1993. We treated the measurements available on this date as baseline $FEV_1$% measurements. 6,028 subjects were excluded due to missing baseline $FEV_1$% measurements (of these, 3,507 were younger than 5.5 years at baseline). After exclusions, 11,254 patients remained in the study cohort; 3,906 (35%) of these patients were observed to die during the study period ending on December 31, 2011. 2,414 subjects were recipients of lung transplantation between 1993 and 2011; we censored these subjects at the time of transplantation.

For each patient, we had baseline demographic and diagnosis data and approximately annual measurements of $FEV_1$%. Counting multiple observations per patient, we included 140,651 total annual records. Of these, 16,846 had missing $FEV_1$% measurements. We imputed missing values by carrying forward the last non-missing value for that patient.

**Risk models**

$FEV_1$% of predicted is a standard measure used in practice to guide recommendation for lung transplantation. We evaluated the following models: (i) a base model containing only $FEV_1$% and (ii) a multivariate model consisting of $FEV_1$%, age, gender, weight, pancreatic sufficiency, *Staphylococcus aureus* infection, and *Burkholderia cepacia* infection.[5] Predictions from Cox models were summarized into a single baseline risk score and a separate time-varying, updated risk score. For the baseline score, we used 10-fold cross-validation to protect against overfitting. For the time-varying score, we used baseline measurements as training data to develop the Cox model and predicted the score at follow-up times using updated values of $FEV_1$%. We added flexibility to both models by using cubic splines to model continuous variables.

**Evaluation of model accuracy**

Diagnostic accuracy: classification error rates

The traditional diagnostic classification problem is based on a binary outcome, typically the presence or absence of disease. Mayer-Hamblett et al[9] assessed prognostic accuracy by defining a yes/no outcome for death within 2 years from baseline. Classification errors include false negatives, i.e. patients who were predicted by the risk score to survive for longer than two years were observed to die within two years, and conversely, false positives, i.e. healthier subjects who did survive beyond two years were predicted to die within two years. Minimizing these errors is equivalent to maximizing the sensitivity (or true positive fraction (TPF)) and specificity (or 1 - false positive fraction (FPF)), respectively. The Receiver Operating Characteristic (ROC) curve is a standard tool that plots TPF versus FPF for all possible risk score cut-offs.[12-16] The ROC curve is commonly summarized using the area under the ROC curve (AUC), ranging from 0.5 to 1.0, which indicate no discrimination to perfect discrimination. The AUC is also a concordance statistic and represents the probability that a randomly chosen subject who dies (case) has a higher marker value than a randomly chosen subject who survives (control).

Time-varying prognostic accuracy

Implicit in the use of traditional diagnostic TPF and FPF are current-status definitions of disease. Since we are interested in a setting where outcome status changes with time, precise definitions are necessary to include event timing in definitions of prognostic error rates. Time-dependent ROC curve methods that extend concepts of sensitivity and specificity and characterize prognostic accuracy for survival outcomes have been proposed in the statistical literature and now widely adopted in practice.[17,18]

      To evaluate both baseline and time-varying measurements, we use the incident-case definitions of Heagerty & Zheng, which are based on a binary classification of the *risk set* at any time $t$.[18] That is, among the patients who are still alive at time $t$, cases are defined as those who die at $t$ and controls as those who survive beyond $t$. The sensitivity and specificity at time $t$ are the error rates in classifying subjects at that time, and can be summarized using a time-dependent ROC curve. The time-dependent AUC, or AUC($t$), is then defined as the area under the time-dependent ROC curve and interpreted as the

probability that a randomly chosen case who dies at time $t$ has a higher marker value than a randomly chosen control who survives beyond time $t$.

These definitions are appropriate for evaluating the performance of a baseline or time-varying marker in the CF lung transplantation setting, as interest lies in identifying patients who are at the highest risk of death in the near future, so that they may be given priority for limited donor organs. The recipient decision may be made at multiple time points as donor organs become available, but is always only applicable to those subjects who remain alive at those times.

We estimated the time-dependent AUC using a simple nonparametric rank-based approach.[19] The idea behind this approach is to compute for each risk set the binary concordance statistic using only the individual case and associated risk-set controls. For a fixed time $t$, we calculate a percentile for each case in the risk set relative to the controls in the risk set. A perfect marker would have the case value greater than 100% of risk set controls. The mean percentile at time $t$ is calculated as the mean of the percentiles for all cases in a window around $t$. The summary curve, AUC($t$), is then estimated as the local average of case percentiles. This nonparametric approach provides both a simple description for marker performance within each risk set and by smoothing individual case percentiles, a final summary curve over time characterizes how accuracy may be changing over time.

Although the AUC is a standard measure of accuracy, it summarizes the sensitivity of a risk score over the entire range of specificities from 0 to 100%. In contrast, clinical decisions are typically made based on a single risk score cut-off that has been shown to perform with high sensitivity and/or high specificity. Therefore, in evaluating a risk score's performance and its impact for treatment decisions, it is important to also assess the sensitivity at a fixed high specificity. Using the above methods, we obtained a summary curve of sensitivity or TPF for a fixed specificity of 95% or FPF of 5%.

Finally, we assessed the time-varying prognostic accuracy of time-varying, or updated, risk scores using an extension of the approach of Saha-Chaudhuri & Heagerty[19] to accommodate time-varying markers (Heagerty, Bansal, Saha-Chaudhuri et al., 2015, unpublished manuscript). At any time $t$, the last measured value of the risk score was used as the current risk prediction.

**Stratification by Risk Group**

We evaluated the performance of annually updated $FEV_1\%$ measurements in subgroups defined by baseline $FEV_1\%$ ($FEV_1\% \leq 30$ versus $FEV_1\% > 30$), baseline age (≤11 years, 12-17 years, and ≥18 years), gender, and F508 genotype.

**RESULTS**

**Cohort characteristics**

Table 1 summarizes the characteristics of the study cohort. Patients who died earlier also tended to be older at baseline, have lower $FEV_1\%$, be in lower weight and height percentiles, be slightly less likely to have *Staphylococcus aureus* infection and slightly more likely to have *Burkholderia cepacia* infection.

Table 1: Summary of baseline (1993) subject characteristics

|  | Overall (n = 11,254) | Died within 1 year (n = 287) | Died 1-5 years (n = 1,077) | Survived at least 5 years (n = 9,146) |
|---|---|---|---|---|
| Age on Dec 31, 1992, Mean (SD) | 18.0 (8.9) | 24.2 (9.6) | 22.0 (9.2) | 16.9 (8.5) |
| Sex, n (%) | | | | |
| • Female | 5,187 (46%) | 144 (50%) | 529 (49%) | 4,185 (46%) |
| • Male | 6,067 (54%) | 143 (50%) | 548 (51%) | 4,961 (54%) |
| Race, n (%) | | | | |
| • White | 10,858 (96%) | 281 (98%) | 1,031 (96%) | 8,819 (96%) |
| • African American | 320 (3%) | 6 (2%) | 40 (4%) | 259 (3%) |
| • Other | 76 (1%) | 0 (0%) | 6 (1%) | 68 (1%) |
| Genotype, n (%) | | | | |
| • ΔF508 homozygous | 4,346 (39%) | 55 (19%) | 281 (26%) | 3,856 (42%) |
| • ΔF508 heterozygous | 3,139 (28%) | 31 (11%) | 165 (15%) | 2,819 (31%) |
| • Other | 883 (8%) | 11 (4%) | 63 (6%) | 742 (8%) |
| • Missing | 2,886 (26%) | 190 (66%) | 568 (53%) | 1,729 (19%) |
| FEV$_1$%, Mean (SD) | 68.5 (28.0) | 32.4 (19.2) | 40.8 (20.1) | 74.5 (25.4) |
| Weight percentile, Mean (SD) | 28.5 (26.6) | 8.7 (16.1) | 12.4 (18.5) | 30.5 (26.6) |
| Height percentile, Mean (SD) | 29.3 (26.3) | 17.5 (23.0) | 19.1 (22.4) | 30.6 (26.5) |
| *Staphylococcus aureus* status, n (%) | | | | |
| • Yes | 3,001 (27%) | 42 (15%) | 242 (22%) | 2,559 (28%) |
| • No | 7,014 (62%) | 234 (82%) | 793 (74%) | 5,468 (60%) |
| • Not cultured | 1,239 (11%) | 11 (4%) | 42 (4%) | 1,119 (12%) |
| *Burkholderia cepacia* status, n (%) | | | | |
| • Yes | 346 (3%) | 33 (11%) | 88 (8%) | 195 (2%) |
| • No | 9,669 (86%) | 243 (85%) | 947 (88%) | 7,832 (86%) |
| • Not cultured | 1,239 (11%) | 11 (4%) | 42 (4%) | 1,119 (12%) |

**Assessment of time-varying prognostic accuracy of base model using baseline and updated measurements**

Figure 1(a) shows the estimated AUC($t$) or mean percentile for FEV$_1$% baseline and annually updated measurements. Not surprisingly, the performance of a baseline FEV$_1$% measurement declines over time, from 0.87 (95% CI (0.84, 0.88)) at baseline (1993) to 0.62 (95% CI (0.59, 0.65)) 20 years later (2012). In contrast, an annually updated FEV$_1$% consistently maintains an AUC of approximately 0.90 over time. Although AUC=0.90 is typically considered to be excellent performance, it does not translate to FEV$_1$% having adequate performance in this setting. 90th percentile means that out of 100 patients on the wait-list, the case marker value is higher than 90 control marker values; however, it also means that 9 other controls will be prioritized ahead of the case. Figure 1(b) shows the TPF for a fixed FPF of 5%. Again, an annually updated FEV$_1$% has better performance than baseline FEV$_1$%; however, a TPF of 45% is likely inadequate for clinical practice.

FEV$_1$% alone does not have satisfactory performance. We also assessed the performance of a multivariate model that combined FEV$_1$% with a number of clinical variables. We note that this model is very similar to Liou et al's model[5], which they proposed as a better predictor of mortality, with the exception that our model excluded diabetes mellitus and number of acute exacerbations, as we did not have data on these variables for this analysis. Figure 2 confirms the findings of Mayer-Hamblett et al[9] and shows that adding clinical variables to the FEV$_1$% base model does not improve performance beyond using FEV$_1$% alone, using baseline or updated measurements.

Finally, an assessment of the performance of annually updated FEV$_1$% measurements compared to those updated every 2 years shows slightly worse performance of the latter in even years when the measurement is a year old. Performance in odd years is exactly the same, since FEV$_1$% is always updated in those years (table available in online supplemental material).

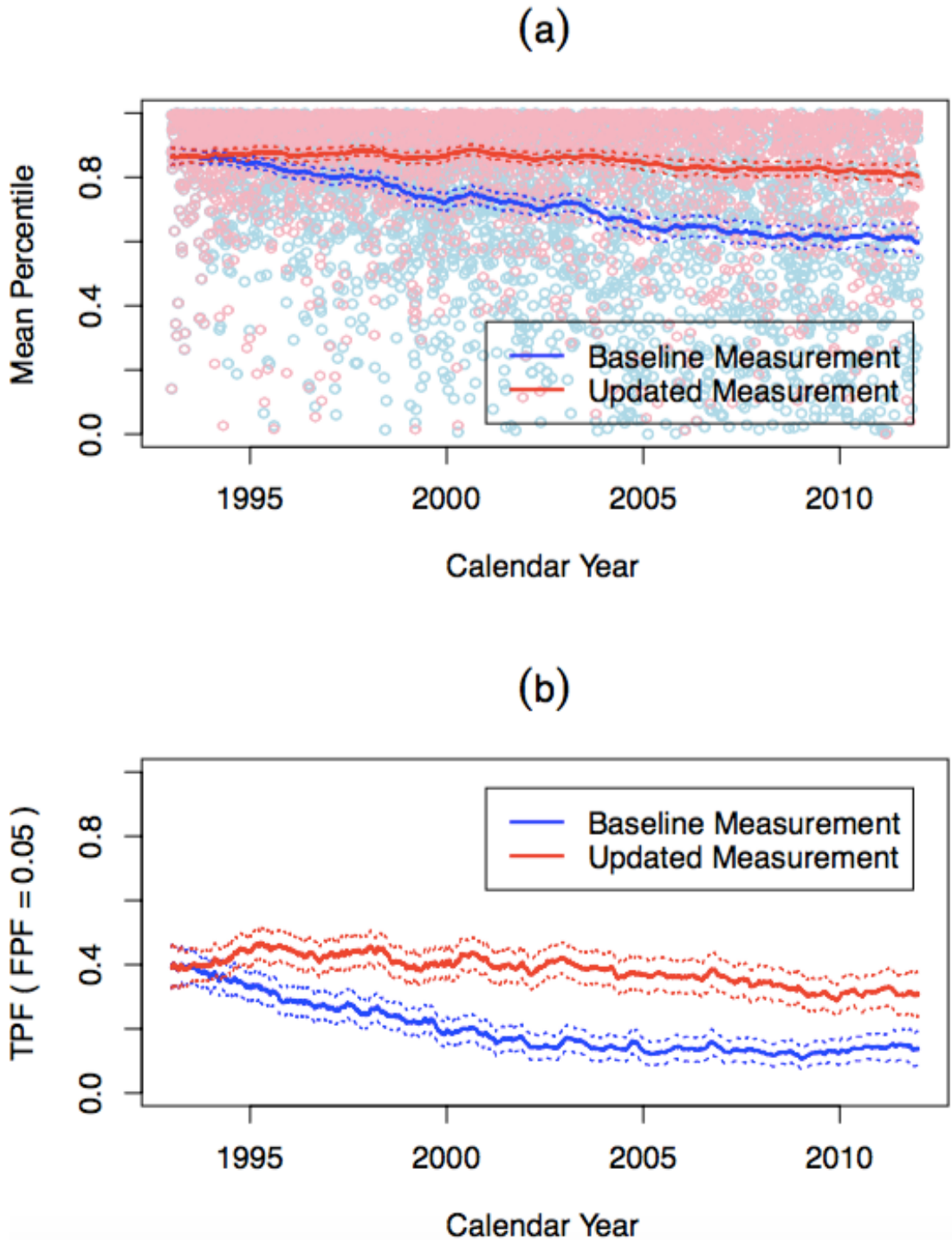

Figure 1: Time-varying performance of baseline versus updated FEV$_1$% measurements, using (a) AUC, and (b) sensitivity at a fixed specificity of 95%. Dotted lines represent 95% confidence bands.

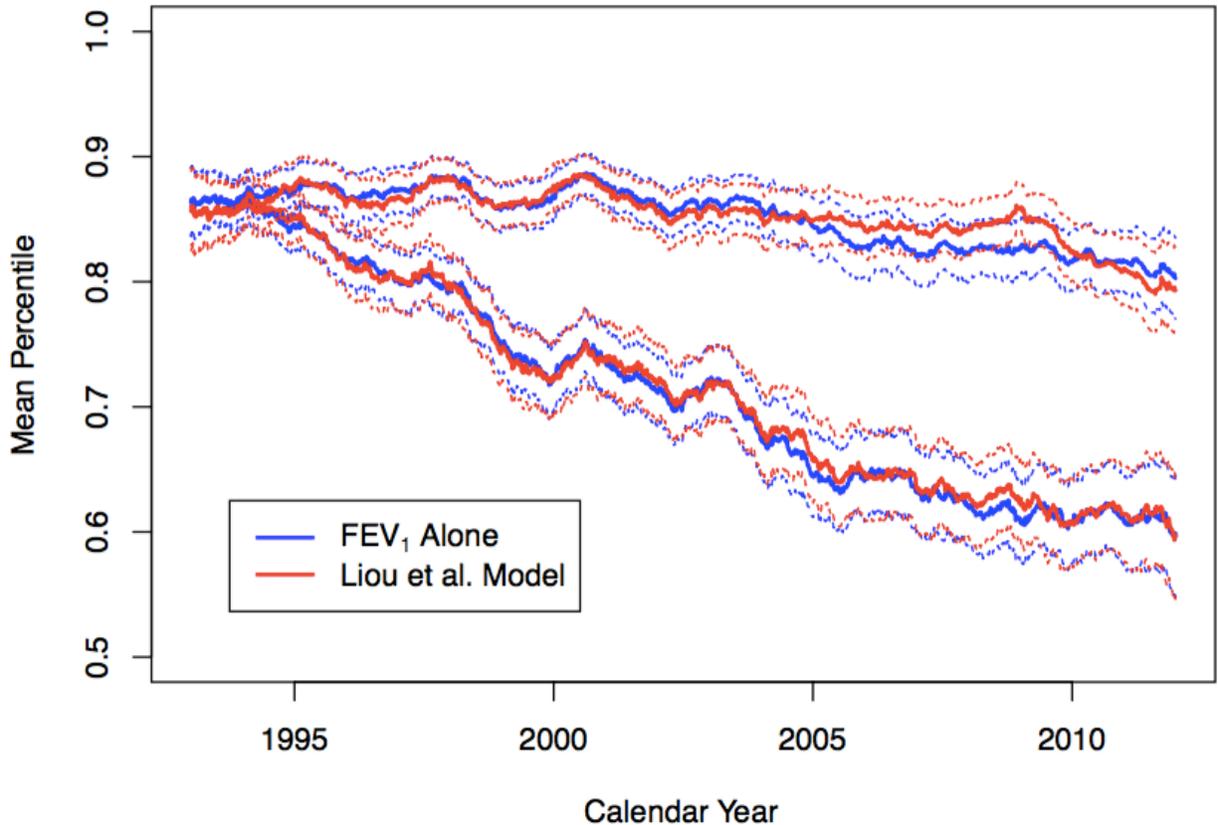

Figure 2: Performance of $FEV_1\%$ alone compared to a multivariate model combining $FEV_1\%$ with other clinical variables, time-varying measurements (top curves) and baseline measurements (bottom curves). Dotted lines represent 95% confidence bands.

**Stratification by risk group**

Figure 3 presents Kaplan-Meier survival curves for subgroups defined by baseline $FEV_1\%$, baseline age, gender, and F508 genotype. We see a large gap in survival probabilities by baseline $FEV_1\%$. Patients with $FEV_1\% \leq 30$ have poor prognosis, with an estimated 5-year survival probability of 35%, compared to 90% in the $FEV_1\%>30$ subgroup. The second panel shows worse survival with increasing age, with estimated 5-year survival probabilities of 95%, 87% and 76% in the $\leq 11$ years, 12-17 years, and $\geq 18$ years subgroups, respectively. Gender and F508 genotype have little impact on survival probabilities. This result regarding F508 genotype is likely due to the low rate of genotyping in this cohort (66% of the patients who died within one year were not genotyped).

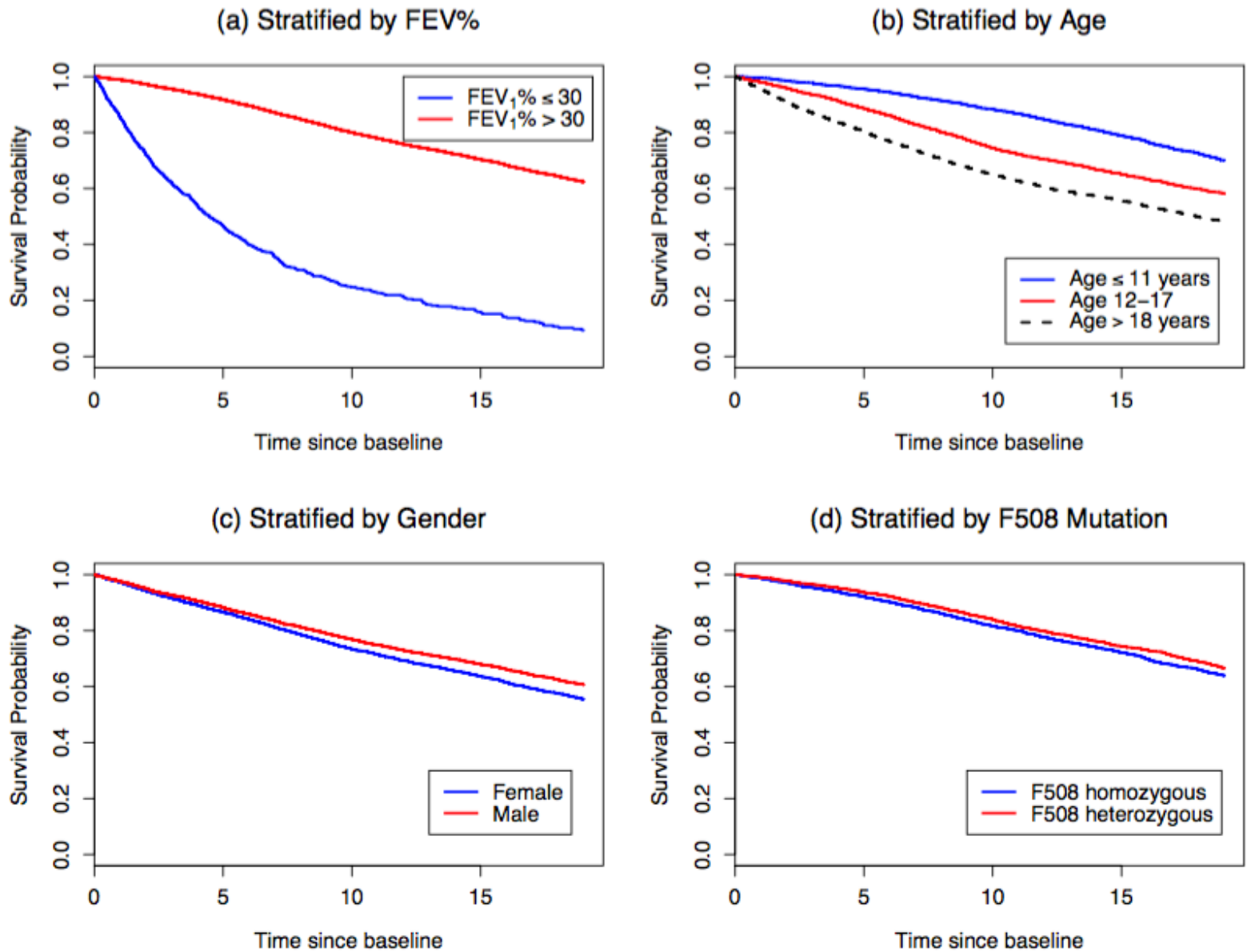

Figure 3: Kaplan-Meier survival curves by subgroup

Figure 4 shows the predictive accuracy of $FEV_1\%$ in the same subgroups. The performance of FEV1% in each subgroup seems to be largely determined by the prognosis in that subgroup. $FEV_1\%$ has poorer predictive performance in subgroups with poorer survival, where it is likely that a patient's prognosis is dominated by other factors that are not captured by $FEV_1\%$.

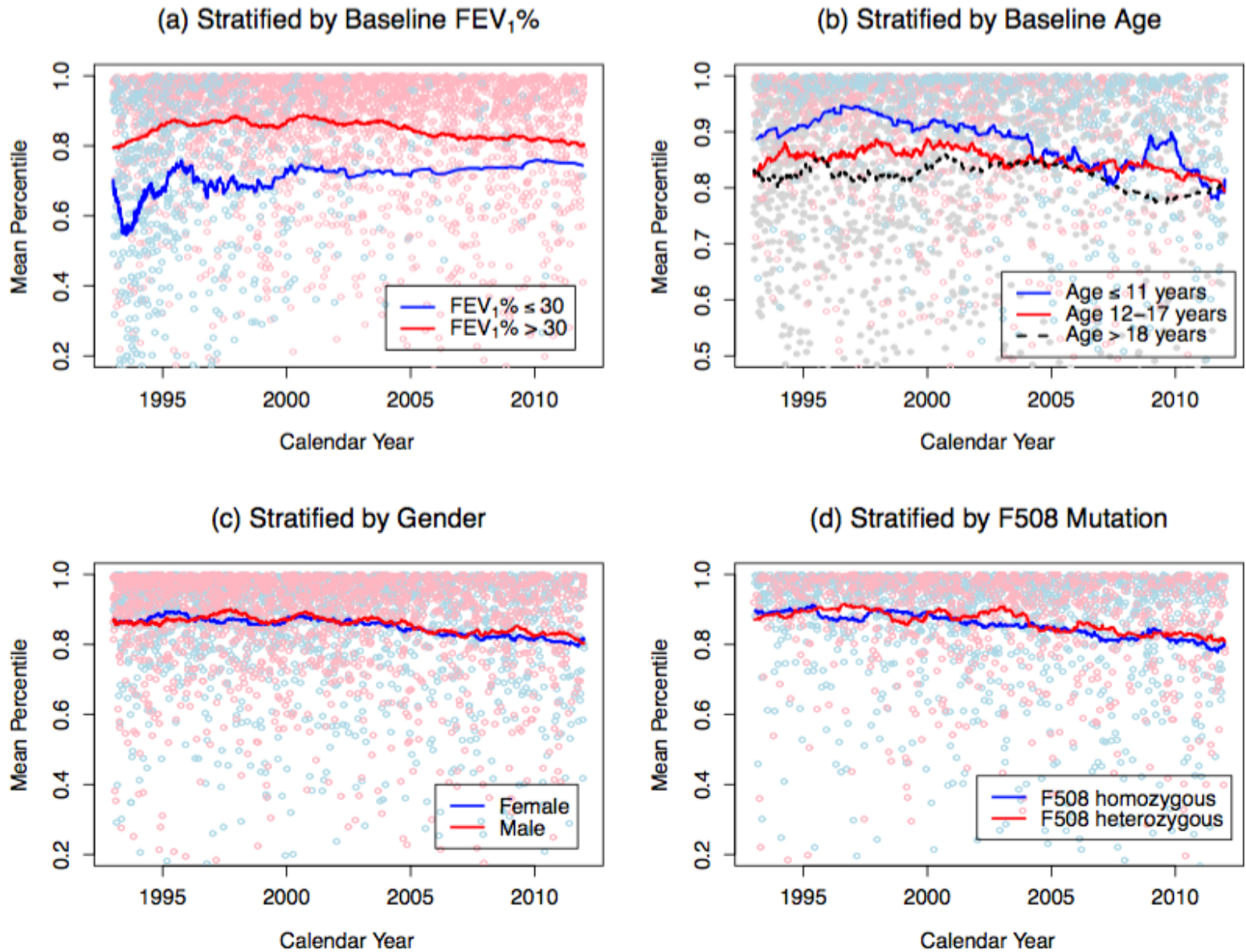

Figure 4: Performance of time-varying FEV$_1$% measurements by subgroups based on: (a) FEV$_1$%, (b) age, (c) sex and (d) F508 mutation

**DISCUSSION**

Commonly used measures of model fit and calibration are insufficient for evaluating risk prediction models in CF when the goal is to prioritize patients for lung transplantation and when a patient's clinical status is changing over time. Therefore, we presented an approach that uses time-dependent classification error rates that can be used to characterize the potential performance of a survival model, while accounting for the time-varying nature of the prediction itself.

In the CF setting, we found that updated measurements of $FEV_1$% have consistent performance over time, whereas the performance of a baseline measurement declines over time. Thus, previously reported estimates of the accuracy of $FEV_1$% alone do not capture its true performance in a clinical setting. It is clear that patient information should be updated over time to maintain classification accuracy; however, it is also evident that neither $FEV_1$% alone nor existing multivariate models are adequate for use in practice.

Being able to evaluate a model's time-varying accuracy may also help guide clinical practice and policy with regards to the frequency of updating patient information. A comparison of 1-year versus 2-year measurements of $FEV_1$%, for example, showed minor differences in performance.

A limitation of the analysis is that it was done in an older cohort, in order to compare performance with Liou et al's model[5]. A more up-to-date cohort could change the results, but it would not take away from the key point – that time-varying approaches are better than the traditional approaches currently used in CF.

In conclusion, we found that using a statistical evaluation approach that is closely tied to the clinical goal of using predicted risk as a score to rank patients as a function of time can significantly change the conclusions drawn about a risk prediction model's performance. As new models are developed, perhaps incorporating novel biomarkers, the proposed approach could be used to accurately assess their predictive ability. As shown, standard methods may underestimate their performance by not capturing how these models will be used dynamically within the clinical setting. We note that our focus here is on risk prediction models, assuming that patients are added to a lung transplantation wait-list based on their expected benefit from transplantation. In practice, any risk prediction should be coupled with assessments of treatment benefit.

**ACKNOWLEDGMENTS**

The authors thank the Cystic Fibrosis Foundation for allowing the use of the registry data.

**CONFLICTS OF INTEREST**

None

**FUNDING**

This project was supported by the NIH (R01-HL072966, UL1TR000423, HL103965, R01-AI101307, P30-DK089507), the PhRMA Foundation, the Cystic Fibrosis Foundation, and the FDA (R01-FD003704).

# REFERENCES

1. United Network for Organ Sharing. 2012 Annual Report of the U.S. Scientific Registry for Transplant Recipients and the Organ Procurement and Transplantation Network. Rockville, MD: U.S. Department of Health and Human Services, Health Resources and Services Administration, Office of Special Programs, Division of Transplantation; 2012.

2. Whitehead B, Helms P, Goodwin M, et al. Heart-lung transplantation for cystic fibrosis. 2: Outcome. *Arch Dis Child* 1991;66:1022-1017.

3. Cystic Fibrosis Foundation. Patient Registry 2013 Annual Data Report. Bethesda, MD: Cystic Fibrosis Foundation; 2013.

4. Flume P. Cystic fibrosis: when to consider lung transplantation? *Chest* 1998;113:1159–61.

5. Liou TG, Adler FR, Fitzsimmons SC, et al. Predictive 5-year survivorship model of cystic fibrosis. *Am J Epidemiol* 2001;153:345–52.

6. Milla CE, Warwick WJ. Risk of death in cystic fibrosis patients with severely compromised lung function. *Chest* 1998;113:1230–4.

7. Hosmer DW, Lemeshow S. Applied logistic regression, 2nd edition. New York, NY: Wiley 2000.

8. Kattan MW. Judging new markers by their ability to improve predictive accuracy. *J Natl Cancer Inst* 2003;95:634–5.

9. Mayer-Hamblett N, Rosenfeld M, Emerson J, et al. Developing cystic fibrosis lung transplant referral criteria using predictors of 2-Year mortality. *Am J Respir Crit Care Med* 2002;166:1550–5.

10. Steyerberg EW, Vickers AJ, Cook NR, et al. Assessing the performance of prediction models: A framework for traditional and novel measures. *Epidemiology* 2010;21:128–38.

11. Aaron SD, Stephenson AL, Cameron DW, et al. A statistical model to predict one-year risk of death in patients with cystic fibrosis. *J Clin Epidemiol* Published Online First: 31 December 2014. doi: 10.1016/j.jclinepi.2014.12.010

12. Green DM, Swets JA. Signal Detection Theory and Psychophysics. New York, NY: John Wiley & Sons 1966.

13. Hanley JA, McNeil BJ. The meaning and use of the area under an ROC curve. *Radiology* 1982;143:29-36.

14. Pepe MS. The Statistical Evaluation of Medical Tests for Classification and Prediction.

Oxford: Oxford University Press 2003.

15. Swets JA, Pickett RM. Evaluation of Diagnostic Systems: Methods From Signal Detection Theory. New York, NY: Academic Press 1982.

16. Metz CF. Basic principles of ROC analysis. *Semin Nucl Med* 1978;8:283–98.

17. Heagerty PJ, Lumley T, Pepe MS. Time-dependent ROC curves for censored survival data and a diagnostic marker. *Biometrics* 2000;56:337–44.

18. Heagerty PJ, Zheng Y. Survival model predictive accuracy and ROC curves. *Biometrics* 2005;61:92–105.

19. Saha-Chaudhuri P, Heagerty PJ. Non-parametric estimation of a time-dependent predictive accuracy curve. *Biostatistics* 2013;14:42–59.

**SUPPLEMENTARY MATERIAL**

Table: One-year average performance of annually updated $FEV_1\%$ measurements compared to $FEV_1\%$ measurements updated every 2 years. Performance in odd years is exactly the same between the two measures, since both are updated in those years.

| Year | Annually updated | Updated every 2 years | Difference |
| --- | --- | --- | --- |
| 1993 – 1994 | 0.87 | 0.87 | 0 |
| 1994 – 1995 | 0.87 | 0.85 | -0.02 |
| 1995 – 1996 | 0.88 | 0.88 | 0 |
| 1996 – 1997 | 0.86 | 0.84 | -0.02 |
| 1997 – 1998 | 0.89 | 0.89 | 0 |
| 1998 – 1999 | 0.87 | 0.85 | -0.02 |
| 1999 – 2000 | 0.85 | 0.85 | 0 |
| 2000 – 2001 | 0.89 | 0.85 | -0.04 |
| 2001 – 2002 | 0.86 | 0.86 | 0 |
| 2002 – 2003 | 0.87 | 0.85 | -0.02 |
| 2003 – 2004 | 0.86 | 0.86 | 0 |
| 2004 – 2005 | 0.85 | 0.82 | -0.03 |
| 2005 – 2006 | 0.84 | 0.84 | 0 |
| 2006 – 2007 | 0.82 | 0.80 | -0.02 |
| 2007 – 2008 | 0.83 | 0.83 | 0 |
| 2008 – 2009 | 0.83 | 0.80 | -0.03 |
| 2009 – 2010 | 0.83 | 0.83 | 0 |
| 2010 – 2011 | 0.82 | 0.79 | -0.03 |
| 2011 – 2012 | 0.81 | 0.81 | 0 |